\documentclass[11pt]{article}
\usepackage{amssymb,amsmath,amsfonts}
\usepackage{graphicx}
\usepackage{graphics}
\usepackage{eepic,epsfig}
\usepackage{cite}

\begin{document}

\title{PROJECT ICARUS: Son of Daedalus \\ Flying Closer to Another Star}
%\author{K. F. Long, M. Fogg, R. K. Obousy, A. Tziolas, A. Mann, R. Osborne, A. Presby \\ Project Icarus Study Group \\ info@icarusinterstellar.org \\ %Presented at the British Interplanetary Society 'Daedalus after 30 years' symposium
%30th September 2009.}

\author{K. F. Long, R. K. Obousy, A. C. Tziolas, A. Mann, R. Osborne, A. Presby, M. Fogg \\ PROJECT ICARUS STUDY GROUP\footnote{info@icarusinterstellar.org} \\ \\  Presented at the British Interplanetary Society 'Daedalus after 30 years' symposium \\(30th September 2009.)}

%, M. Fogg, R. K. Obousy, A. Tziolas, A. Mann, R. Osborne, A. Presby
%Project Icarus Study Group\\
%info@icarusinterstellar.org\\
%Presented at the British Interplanetary Society 'Daedalus after 30 years' symposium
%30th September 2009.\\}

%\textit{$^1$ }\\

\maketitle

\section{Abstract}

\label{sec:Introd}

During the 1970s members of the British Interplanetary Society embarked on a landmark theoretical engineering design study to send a probe to Barnard's star. Project Daedalus was a two-stage vehicle employing electron beam driven inertial confinement fusion engines to reach its target destination. This paper sets out the proposal for a successor interstellar design study called Project Icarus. This is an attempt to redesign the Daedalus vehicle with similar terms of reference. The aim of this study is to evolve an improved engineering design and move us closer to achieving interstellar exploration. Although this paper does not discuss prematurely what design modification are likely to occur some indications are given from the nature of the discussions. This paper is a submission of the Project Icarus Study Group.

\section{Introduction}

Sending robotic probes to solar systems other than our own is a vastly different technical challenge when compared against interplanetary exploration. The nearest stars are over four light years away or a distance of $\approx272,000$ Astronomical Units (AU). For this reason, history has shown a reluctance to accept the prospects for interstellar travel as realistic. Since 1999 however, astronomical observations of other stars have identified over 400 extra-solar planets, primarily using the radial velocity method \cite{Schneider,Butler,Udry}. This has led to the growing possibility that within decades astronomers may identify a habitable world which could, in theory, be one day explored by humans. The technical challenge remains however - even when you know where you want to go how do you get there?

There have been numerous theoretical studies into different propulsion schemes one may use to cross the vast distance of space. A quick review of the various concepts will immediately show that many of those schemes are ruled out. This includes chemical, electric and nuclear fission \cite{Long1}. This is largely due to either a lack of specific impulse or a lack of thrust to give high exhaust velocity. A high exhaust velocity would be necessary to reach the speeds that are required to travel to another star within the time span of a human lifetime - one of the Terms of Reference of the Icarus mission to be discussed later in this paper. Fuels with comparatively low levels of energy available, for liberation in the form of thrust, would also require a high mass ratio placing constraints on how far the spacecraft can go. \cite{Mallove}

Although there are exotic concepts for interstellar travel such as warp drive and wormholes, these are currently considered to be purely speculative \cite{Long2,Obousy} . Concepts such as antimatter have been investigated but the technology is currently too immature to harness for a spacecraft \cite{Forward1}. Employing the energy from the Sun in solar sail driven vehicles is credible but has the problem that solar intensity reduces inversely with the distance squared. This can be compensated for by using large collimated laser beams \cite{Forward2}, but this technology has not been demonstrated for such an application. In the search for fuels which are energetic, provide for low mass ratios, low Thrust/weight ratio for high exhaust velocities, designers are led to consider nuclear pulse engines, in particular with a fusion based fuel.

The idea of using nuclear pulse propulsion was first proposed by Stanislaw Ulam in 1947 and subsequently resulted in an engineering study led by Ted Taylor \cite{Taylor}. This was to employ nuclear pulse technology to propel a vehicle by capturing the blast products of a nuclear explosion on the rear of a giant pusher plate, transferring momentum to the spacecraft and its occupants, cushioned from the blast by several shock absorbers. This eventually led to the Project Orion engineering study with around \$11 million being spent over a period of 7 years. The design was for a mission which would take only $approx$140 years to reach Alpha Centauri with 1 `unit of propellant' being exploded every 3 seconds to push the spacecraft at 1g acceleration to $\approx3\%$ of light speed [11]. However, with the arrival of various test ban treaties external nuclear pulse was shelved to the archives. Despite this, Project Orion remains the first comprehensive interstellar spacecraft design which was credible.

In 1971 Friedwardt Winterberg explored the concept of using Marx generators to power electron particle beams \cite{Winterberg}. This idea was picked up by members of the British Interplanetary Society (BIS) who were considering embarking on an engineering design study to demonstrate that interstellar travel was, at least, possible with current, or near future, technology. Then in January 1973 members of the BIS first met to discuss the challenges of interstellar propulsion and the idea of Project Daedalus was born. Led by Alan Bond, Tony Martin and Bob Parkinson, members came together to create what has become one of the most comprehensive interstellar engineering studies ever undertaken. Project Daedalus had three stated guidelines:

\begin{enumerate}
\item The spacecraft must use current or near future technology.
\item	The spacecraft must reach its destination within a human lifetime.
\item	The spacecraft must be designed to allow for a variety of target stars.
\end{enumerate}

The members of the Daedalus study group were all volunteers but with a solid
knowledge of engineering and science. The final design was published in 1978 \cite{Bond}
and was a two-stage spacecraft nearly $200$m in length powered by electron driven
D/He$^3$ fusion reactions, eventually accelerating up to $12$\% of light speed to arrive at
its target destination Barnard's star 5.9 light years away in under $50$ years. The
Daedalus first stage had a structural mass of $1,690$ tonnes with $46,000$ tonnes of
propellant in six tanks and it would burn for 2.05 years before jettisoning it. The
second stage had a structural mass of 980 tonnes with 4,000 tonnes of propellant in
four smaller tanks and would burn for 1.76 years, the tanks all being jettisoned prior to
reaching the destination. Both stages would achieve an exhaust velocity of around
10,000 km/s from the detonation of the D/He$^3$ pellets at a frequency of 250 Hz. The
second stage carried a $450$ tonne scientific payload with 18 probes to be used to study
the target solar system planets. One of the interesting problems that Daedalus had to
deal with is the transmission of a radio signal over vast distances back to earth. The
team proposed the unique solution of using the 2nd stage parabolic reaction chamber to
achieve high antenna gain. It is too early in the design process (pre-concept stage) to
identify how Project Icarus may do things differently, giving the changes in
technology over the three decades. However, at this stage we can speculate that likely
design modifications will be in the areas of electronics, computing science probes,
pellet ignition driver, the propellant and its acquisition. Some of the issues will be
briefly discussed in this paper. 

Project Daedalus also addressed the fundamental question of the whereabouts of intelligent life in the universe given that sufficient time had existed for sun-like stars to evolve with potentially habitable planets, despite the large distances between the stars, a problem termed the Fermi Paradox. Project Daedalus showed that if humans could envisage a vehicle capable of crossing those distances in a reasonable amount of time, then later more efficient ways of doing it would be found. The project showed that interstellar flight was most likely possible and more sophisticated thinking about the absence of intelligent visitors would be required \cite{Parkinson}. 

In addition to Project Daedalus, a number of alternative fusion based propulsion studies have also been performed. One of these was Project Vista \cite{Orth} in the 1980's which proposed using D/T fusion reactions driven by a $\approx 5$ MJ laser beam to execute a mission to Mars within 6 months or any planet in the solar system within 7 years. Vista would employ a technique known as fast ignition to achieve an energy gain compared to the input energy of $>1000$ and a specific impulse of $\approx10^4 {\rm s}$. Another study in the 1980's was Project Longshot \cite{Beals} which proposed using D/He$^3$ reactions driven by either laser or electron beams to accelerate a mission to Alpha Centauri within a century by reaching a speed of $5\%$ of light and by having a specific impulse of $10^6 {\rm s}$. The He$^3$ was to be manufactured on earth using particle accelerators which bombard a lithium blanket. Alternatively, it was to be captured from the solar wind or mined en route from the atmosphere of Jupiter. 

The most distant robotic pioneers sent out of the solar system are the Pioneer and Voyager spacecraft, launched in the 1970's. Travelling at slow speeds of only a few AU per year, these spacecraft would take tens of thousands of years to reach the distance of nearest stars if they continued in their current direction. There is a large technological gap which needs to be breached, if interstellar exploration is to become a reality. How do we bridge this gap? One answer is to perform a focussed design study for a particular propulsion engine with the end objective of increasing the Technology Readiness Level or TRL of the design type. These describe a range of technological maturities from conjecture through to a tested application \cite{Mankins}.  Hence one of the main purposes behind Project Icarus and the decision to base the engine design on mainly fusion technology.

\section{Fusion Research}

It has been known for sometime that the Sun is powered by nuclear fusion reactions, a process sustained over billions of years by its large gravity field. When this was first realized it soon became obvious that the potential existed to generate the same fusion reactions on Earth for use in electrical power generation. A fusion reactor would be much safer than conventional fission reactors, due to much lower levels of radioactive contamination. Typical fusion reactions include Hydrogen and Helium isotopes such as D(T, $\alpha$)n; D(D, He$^3$)n; D(D, T)p and D(T, $\alpha$)p.

The ongoing fusion programme on Earth has been a constant reminder that the application of fusion energy to space propulsion is a likely prospect \cite{Schulze}. Historically, scientists have been disappointed with the lack of progress in producing a sustained fusion reaction in the laboratory. The key requirement to attaining ignition is the fusion triple product which states that the product of the temperature, particle density and plasma confinement time must be  $\geq 10^{21}{\rm m}^{-3} {\rm sKeV}$. This is achieved by two main routes. Firstly, one can utilise a low density $(\approx 10^{14}{\rm cm}^{-3})$ of particles for a long time ($\approx$ seconds) and confine the plasma using magnetic fields - this is the Magnetic Confinement Fusion (MCF) concept. The other route is to utilise a high density $(\approx10^{23} {\rm cm}^{-3})$ of particles but for a short time ($\approx{\rm ns}$) such as using laser beams which deliver $\approx1  {\rm MJ}$ to the target - this is the Inertial Confinement Fusion (ICF) concept, so named because the inertia of the material itself is being used to confine the plasma. This was the route adopted for the Daedalus study. 

Typically a plot of the reaction average cross section as a function of plasma temperature will show that D/T reactions will ignite at the lowest temperature, followed by D/D and D/He3. A technical issue for any fusion based space propulsion is the choice of fuel to use. D/T has the lowest ignition barrier due to the high ratio of neutrons (binds nuclei via strong force) to protons (repels nuclei apart via Coulomb force). There is also another approach to fusion ignition known as Inertial Electrostatic Confinement Fusion (IECF) which works on the basis of accelerating charged particles radially inwards using an electrostatic field. This is the basis of so called table top fusion reactors but will not be discussed further here. Instead, we focus on the confinement method as used in the Daedalus engine design.

In ICF, typically a pellet would be a D/T fuel surrounded by Beryllium with a diameter of around 2mm. For reference the pellets proposed for Daedalus were much larger of order 2 to 4 ${\rm cm}$ diameter requiring $\approx3\times 10^{10}$ pellets which equated to an annual production rate of $~1000$ per second. The engine would produce $\approx1.5 \times 10^{24}$ neutrons per second for the first stage and $\approx 1.1\times 10^{23}$ neutrons per second for the second stage \cite{Bond}. The confinement of a fusion pellet is complicated by the interaction of the beam \cite{Pfalzner}. In the case of a laser beam, mass is initially lost to the surface via ablation and sets up a coronal plasma layer. The laser beam can then only penetrate so far, defined by a critical density surface, and instead it is absorbed into the electrons, mainly by collisional inverse Bremsstrahlung, collisionless resonance absorption and parametric instabilities (excitation of plasma and electromagnetic waves). Then the electrons transport the energy inwards to transfer their energy into the thermal energy of the compression. Convergent shocks with an implosion velocity of $\approx 4\times 10^7  {\rm cm/s}$ then compress the fusion fuel up to $100$ times the density of lead producing a high temperature $10^8 {\rm K}$ hot spot at the centre of the pellet. When the conditions are right, the fusion reactions will begin depositing alpha particles into the surrounding high density but cooler fuel, and a fusion burn begins outwards defining the point of ignition. Provided this burn up occurs faster than the time for the pellet to expand under its own pressure, something like $\approx20 {\rm MJ}$ of energy would be liberated \cite{Nuckolls,Lindl}. The objective is to achieve energy gain; i.e. more energy would be released from the fusion reaction than the driver energy required to achieve the compression and burn. 

To achieve this process in a spacecraft is no small challenge. The spacecraft must have the apparatus of a driver beam (laser, electron) and some way of accelerating the pellets into the target chambers. To achieve this, the Daedalus design coated each pellet with a superconducting magnetic shell and then accelerated the pellets into the chamber with an electromagnetic gun. Electron beams would then confine each pellet for a fraction of a second until ignition conditions were achieved. The Daedalus design required 250 pellets to be detonated per second and then redirected the reaction products via a magnetic nozzle for thrust \cite{Bond}. This entire operation is a technical challenge in itself even for a reactor stationed on earth. 

In recent years, the fusion energy program is moving ahead as demonstrated with the arrival of several important facilities to achieve fusion ignition in the laboratory. The Joint European Torus (JET) began operation in 1983 and eventually produced around 40MW of fusion power for $\approx 1$ second \cite{Pick}. This has led directly to the design of the International Thermonuclear Experimental Reactor (ITER), currently under construction in France and due to begin operation in 2018 \cite{ITER}. Once operational it should produce around $500 {\rm MW}$ of fusion power for $\approx 1000$ seconds and produce a gain of $ > 10$. This is a true technology demonstrator for a workable nuclear fusion generator plant.  Both JET and ITER rely upon the MCF scheme to achieve ignition. 

There have also been numerous laser facilities throughout the world, culminating in the creation of the US National Ignition Facility (NIF) which began operation in 2009 \cite{Hammel}. The NIF uses 192 beams to deliver $1.8 {\rm MJ}$ to an ICF pellet producing $100 {\rm MJ}$ of fusion power for around $1 {\rm ns}$. The French are also building a similar facility known as Laser MegaJoule (LMJ) \cite{Fleurot}. Both NIF and LMJ relies upon the ICF scheme to achieve ignition and would produce a gain of $ >10 $. A further proposal for the future is a facility called High Power Laser for Energy Research (HiPER) \cite{Atzeni}. The project plans to use the fast ignition technique to achieve a high gain of $>100$. Fast ignition is a method of separating the compression and ignition phases of the ICF pellet implosion using a combination of long pulse and short pulse laser beams. There are exciting times ahead in fusion energy research and it is quite possible that the long standing ambition of harnessing the power of the Sun in a laboratory on earth will be achieved within years. With these rapid developments, it is a good time to review how this will affect the potential for a fusion based spacecraft on a mission to the nearest stars. 

\section{Extrasolar Planet Observations}

Of the more than 400 extrasolar planets so far discovered most are gas giants as these are a lot easier to detect using the radial velocity method \cite{Schneider, Butler}. Approximately, $7\%$ of F/G/K class stars and $1\%$ of M class stars surveyed possess a detectable giant planet - a proportion that is likely to increase as observational techniques and baselines improve. The first space based survey to search for extra-solar planets was a French mission called CoRoT launched in 2007. The spacecraft has been successful in detecting several planetary transits around other stars, including the first super-Earth planet with a measured radius \cite{Leger}. The Kepler spacecraft incorporates one of the largest space cameras covering 105 square degrees of sky allowing it to examine around 100,000 stars between 150-2500 light years away in the area of the constellation Cygnus. Kepler will determine if any of them have earth like worlds (particularly in the habitable zone) using the transit technique, which searches for changes in the brightness of a star as a planet crosses it. Kepler represents a significant advance on earth based observatories that have had to deal with optical limitations imposed by the presence of the atmosphere. Kepler will also be an advance on CoRoT and should detect planets with much larger orbital periods \cite{YEE}. Other proposals on the drawing board include NASA's Terrestrial Planet Finder mission \cite{Lawson} and the ESA's Darwin mission \cite{Cockell}. Such space-based observatories, should anything like them eventually be built, will attempt the direct imaging of terrestrial planets around stars in the solar neighbourhood and the analysis of their spectra for atmospheric molecules and biomarkers.

The Daedalus project proposed visiting Barnard's star, situated 5.9 light years away, within a mission duration of under 50 years at a probe cruise velocity of $~12\%$ of light \cite{Bond}. Part of the reason for choosing this particular target system was because of a supposed astrometric detection of a Jovian type planet orbiting Barnard's star - claims that are now considered to be erroneous \cite{Gatewood}. A mission design at the same cruise velocity which aims for a century or less duration restricts target selection to within 12 light years from the Sun - a tiny distance in comparison to the size of the Milky Way, but still containing a choice of 20 star systems, including 6 binary and 3 triple star systems.

At the present time, the closest system thought to host an exoplanet is that of the young $\approx600$ Myr old star Epsilon Eridani, $\approx10.7$ light years distant. Radial velocity observations suggest the existence of a giant planet of $\approx1.55$ Jupiter masses in an eccentric orbit centred on a semi-major axis of $\approx 3.39 {\rm AU}$ \cite{Hatzes}. Infrared observations have also detected the presence of dust rings in the system with a structure that hints at collisional evolution within   Eridani analogues of asteroids and Kuiper belts. Sculpting of the inner edge of the outermost dust belt is suggestive of the presence of another planet of $\approx 30$ Earth masses at $\approx$40 {\rm AU} \cite{Quillen}, whilst a third planet may be implied by the structure of the inner belts within the system \cite{Backman}. Although the detection of non-transiting Earth-mass planets is still beyond the state of the art, it has been estimated that the habitable zone of the   Eridani system (sited between $\approx0.48 - 0.93 {\rm AU}$ from the central star) exhibits partial dynamical stability in the sense that a terrestrial planet could have survived within the inner part of the zone with a stable orbit over the lifetime of the system \cite{Jones}.

Another interesting system within the solar neighbourhood is that of $\tau$ Ceti, which is a
single G-class star,$\approx$11.9 LY distant, and a little less massive than the Sun. No giant
exoplanets have been discovered orbiting $\tau$ Ceti, but infrared astronomy has detected
a prominent debris disc around the star extending outwards to $\approx $55 AU \cite{Greaves}. The dust
in this disc is thought to have been generated via collisions within a Kuiper belt like
structure and is evidence that bodies at least as large as comets were able to form in
the $\tau$ Ceti system. It is possible that $\tau$ Ceti, which is only about a third as metal rich as
the Sun, and roughly twice as old, did not originally have a protoplanetary disc
massive enough to form giant planets but may nevertheless be accompanied by a suite
of terrestrial planets as yet undetected.

By the time it becomes possible to construct and launch interstellar probes, perhaps by the second half of this century, it is most likely that the capability of astronomical observations from the solar system will be greatly advanced over that of the present day and near future. Observations may be sufficiently resolved to reveal significant detail about the presence and basic parameters of planetary systems around all the potential target stars in the solar neighbourhood. Modern planet formation theory leads to the expectation that planetary systems should be commonplace. Even if giant planets do not have time to form within a particular protostar's circumstellar disk phase, there seem to be few obstacles in the way of terrestrial planets forming around most stars whose systems are not subject to violent dynamical perturbations during their primordial epoch. Even systems previously considered to be unlikely candidates for hosting planetary systems may surprise us. Exoplanets have been discovered in binary star systems \cite{Eggenberger} and numerical simulations suggest that a set of terrestrial planets could have formed around both components of the G/K class binary at the heart of the Alpha Centauri system, including within each star's habitable zone. \cite{Barbieri,Quintana}. These are the closest sun-like stars to the solar system at 4.37 light years distant. Even the opinion that M-dwarf stars are incapable of hosting habitable planets is now being reappraised \cite{Scalo} and there are 19 individual M class stars within 12 light years of the solar system.

Hence, with the exception of the case of   Eridani, we know little more about what awaits discovery around stars in the immediate solar neighbourhood than the original Daedalus team knew in 1978. However, this ignorance will dissipate well in advance of any future time when it actually becomes possible to launch an interstellar flight and one can be confident that numerous tempting mission targets will have emerged.

\section{Project Icarus}

Project Icarus: son of Daedalus - flying closer to another star. This is the vision of this theoretical design study to re-examine the engineering solution and fundamental assumptions behind Project Daedalus. Like Daedalus, the intention is for Icarus to use fusion based engines and to quote Alan Bond from the recent BIS symposium ``now we are addressing the universe on its own terms''. This was in reference to the utility of using fusion power to visit the stars - which are themselves fusion powered.

Icarus was a character from ancient Greek mythology. In an attempt to escape the labyrinth prison of King Minos, his father Daedalus fashioned a pair of wings for both himself and his son made of feathers and wax. But Icarus soared through the sky joyfully and flew too close to the Sun melting the wax on his wings. He fell into the sea and died after having `touched' the sky. Project Icarus aims to `touch' the stars and escape from the bounds of mother Earth. In the introduction to the Daedalus study report Alan Bond states that ``it is hoped that these `cunningly wrought' designs of Daedalus will be tested by modern day equivalents of Icarus, who will hopefully survive to suggest better methods and techniques which will work where those of Daedalus may fail, and that the results of this study will bring the day when mankind will reach out to the stars a step nearer'' \cite{Bond}. So in essence, the naming of a successor project as Icarus was suggested by the original study group. 

Project Icarus is a Tau Zero Foundation initiative in collaboration with The British Interplanetary Society and so represents a true collaboration of international volunteers all sharing in the vision of a human presence in space in the coming centuries. This gives the project a strong support base and a large intellectual resource. This is represented by the Project Icarus logo shown in Figure 1. The black wings of Icarus reach up to touch the distant star, whilst the feathers on the tips of the wings represents the technology (squares) and spirit or theoretical ideas (circles) of how this vision will be achieved. 

The purpose of Project Icarus is as follows:
\begin{enumerate}
\item To design a credible interstellar probe that is a concept design for a potential mission in the coming centuries. 
\item	To allow a direct technology comparison with Daedalus and provide an assessment of the maturity of fusion based space propulsion for future precursor missions.
\item	To generate greater interest in the real term prospects for interstellar precursor missions that are based on credible science. 
\item	To motivate a new generation of scientists to be interested in designing space missions that go beyond our solar system. 
\end{enumerate}
The Terms of Reference (ToR) for Project Icarus essentially represent the initial design requirements and are as follows:
\begin{enumerate}
\item	To design an unmanned probe that is capable of delivering useful scientific data about the target star, associated planetary bodies, solar environment and the interstellar medium. 
\item	The spacecraft must use current or near future technology and be designed to be launched as soon as is credibly determined. 
\item	The spacecraft must reach its stellar destination within as fast a time as possible, not exceeding a century and ideally much sooner.
\item	The spacecraft must be designed to allow for a variety of target stars. 
\item	The spacecraft propulsion must be mainly fusion based (i.e. Daedalus).
\item	The spacecraft mission must be designed so as to allow some deceleration for increased encounter time at the destination. 
\end{enumerate}
There are also several key watchwords for Project Icarus to ensure that all design solutions are appropriate. The final design must be a CREDIBLE proposal and not based upon speculative physics. It must be a PRACTICAL design. It must be derived using accepted natural laws and using SCIENTIFIC methods which are supported by experiments. It must be based upon only NEAR-FUTURE technology as determined by simple linear extrapolation of current technologies. The team must produce an ENGINEERED design as though the vehicle were close to flight readiness, to ensure that approximations and margins are appropriate. There is also the addition of a RELIABILITY watchword, to ensure that upon arrival all of the scientific instruments will function and this implies redundancy. There is also the addition of a project scope as follows:
`{\it The required milestones should be defined in order to get to a potential launch of such a mission. This should include a credible design, mission profile, key development steps and other aspects as considered appropriate.}'

\begin{figure}
\begin{center}
\includegraphics[width=7.029cm, height=7.129cm]{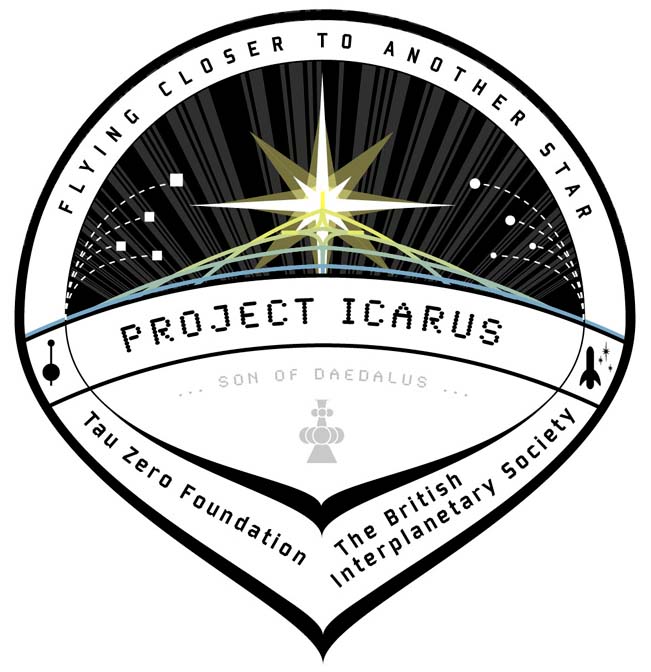}
\caption{Project Icarus official logo (Alexandre Szames/antigravite).} \label{fig1}
\end{center}
\end{figure}

During the project many aspects of the original Daedalus design will be re-examined. This includes the choice of fuel which for Daedalus was D/He$^3$. Alternative options include D/T, D/D, Li/p, B/p although the non-use of D/He$^3$ is likely to dramatically impact the potential specific impulse for the mission. A source of fuel other than the gas giant Jupiter (the choice of the Daedalus designers) should be considered. Another author \cite{Lewis} has re-examined the question of gas giant mining for He$^3$, specifically for Jupiter, Saturn, Neptune and Uranus. The composition for all these planets is expected to be about 45 parts per million (similar to the Sun). There are two basic problems with gas giant mining however. The first is overcoming the huge gravity wells. The escape velocities for the various gas giants are $21.3 {\rm km/s}$ (Uranus), $23.5 {\rm km/s}$ (Neptune), $35.5 {\rm km/s}$ (Saturn) and $59.5 {\rm km/s}$ (Jupiter). 

The second big issue for He$^3$ mining is extraction from the atmospheres of the gas giants. The problem of fuel extraction could be solved, for example, by launching a spacecraft to Uranus orbit, which would arrive in approximately 7 years and then  by dropping a probe into the atmosphere.  An inflatable gas bag would then be used to move through the vertical atmosphere until a few atmospheres of pressure had been reached. Neutral buoyancy would then be maintained. Small fission reactors would be used to power pumps and any refrigeration for the liquefaction and extraction of the He$^3$ by separating it from the He$^4$. Any other excess gasses can be either dumped or used to cool the systems on board. The equipment would then be jettisoned and the fuel tank essentially launched back into orbit to rendezvous with the awaiting vehicle. Alternatively, the material can be launched on a highly elliptical orbit and returned to either Earth or Mars after several years. Historically distance would also have been perceived to be an issue, but with today's technology the entire outer solar system is reachable. From the Earth the gas giants are located at $5.2 {\rm AU}$ (Jupiter), $9.5 {\rm AU}$ (Saturn), $19.2 {\rm AU}$ (Uranus) and $30.1 {\rm AU}$ (Neptune). Overall, all of the gas giants have great potential, although it is worth noting that Uranus is approximately half the distance of Neptune and has a much lower gravity well. 

Daedalus used ICF driven by an electron beam. One of the key elements to achieving nuclear fusion is the attainment of temperatures sufficient to overcome the Coulombic barrier that prevent nuclei from fusing. One exciting possibility for achieving the necessary temperatures is in the use of antimatter. Antimatter was first predicted to exist in 1928 after the British Physicist Paul Dirac realized that the relativistic version of the Schrodinger equation allowed for the possibility of anti-electrons. Since then, antimatter has been produced, albeit in minute quantities, artificially in labs. In Antimatter Catalyzed Micro Fusion (ACMF) a beam of antiprotons is injected into a fusion fuel. The negatively charged antiprotons effectively 'screen' the Coulomb potential repelling the nuclei, reducing the temperature of fusion relative to conventional thermonuclear plasmas \cite{Kammash, Perkins, Frolov}. As well as the more traditional fusion cycles, more exotic aneutronic fuels, for example P-B$^{11}$, have also been considered possible candidates for the fusion fuel, \cite{Kammash1} with reported advantages in specific impulse due to the reaction products being solely charged particles which can be more readily directed for thrust.

One of the biggest assumptions of Project Daedalus was the use of a flyby probe. Project Icarus will re-examine this assumption, and the terms of reference for the study stipulate that deceleration of the spacecraft at the target destination is a mission requirement. One option could be the use of mini-probes which are separated from the main bus to be decelerated into the target system whilst the main spacecraft carries on in its trajectory. Another option could be achieved by using a magnetic sail otherwise known as a MagSail. This uses a magnetic field to deflect charged particles contained within the solar wind emitted from the destination star, which decelerates the spacecraft. \cite{Zubrin}. 

The mission outline of any spacecraft is meticulously coordinated through computers which carry all of the responsibilities associated with the spacecraft's nominal operation. The navigation and attitude control of the spacecraft, communications and science packages are all carefully timed and executed through the program which maintains the mission profile. In traditional spacecraft missions, within the solar system, these commands are carefully monitored by a ground station and corrections are often made when there are changes to the mission parameters. Interstellar missions however, do not have this luxury.

The primary obstacle that needs to be overcome is the very long flight times, during which the mission profile is likely to change several times. Thus all interstellar missions, despite whichever mission outline they may be designed around, effectively have open mission profiles. The on-board computers are for all given purposes, both unmanned and unserviceable and must therefore employ advanced decision trees to make them as redundant as possible. A sophisticated decision making program, a desirable trait of any spacecraft, here becomes a necessity. In fact, the computer in this case must closely resemble an artificially intelligent design.

Any deep space mission will begin inside our solar system, while the trajectory burns and fuel acquisition stages are executed. During these early stages, conventional power systems such as the highly successful and proven photovoltaic technologies can be used. The power generated is proportional to the magnitude of the solar flux which scales as the inverse squared distance to the Sun. However, the efficiency of the photovoltaic panels themselves increases at lower temperatures, which leads to a final effective energy gain which is approximately proportional to $I_e/R^{1.5}$, where $I_e$ is the intensity at earth and $R$ is the orbital radius. The relative power generated, as shown in Figure 2 is only really effective within the range of the terrestrial planets, and so other methods of generating energy should be used \cite{Patel}. 

\begin{figure}
\begin{center}
\includegraphics[width=13.875cm, height=8.425cm]{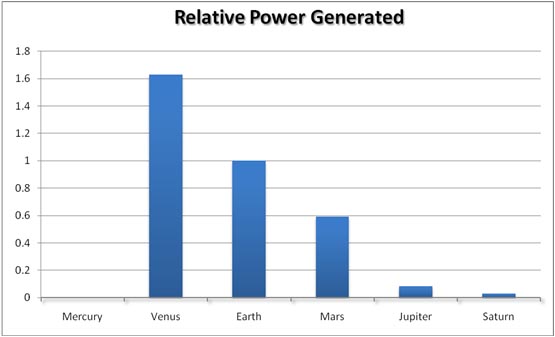}
\caption{Relative power generated by current space grade photovoltaic cells, scaled by planetary orbital radius. Note that at the extremely high temperatures found at Mercury, solar cells are completely inoperable.} \label{fig2}
\end{center}
\end{figure}

Other methods of power generation such as the use of space tethers, where a cable is used as a source of power while it moves through a planet's magnetic field, could only be useful to the Icarus during its fuel acquisition phases, if a planetary mining scenario similar to that of Daedalus is selected. Fuel cell technologies, using electrochemical reactions have seen large advances in both cell lifetime and power output. Several design scenarios exist which use reversible reactions, leading to regenerative cells, which can be used as reliable backups when there are main power interrupts on the spacecraft. 

Flywheels store rotational kinetic energy and are extremely efficient ways of storing energy. NASA designed its G2 Flywheel Module to be a low-cost modular test bed for flywheel system integration, using magnetically levitated carbon-fibre wheels with permanent magnet motors. The latest designs show that all of the system losses can be addressed by conductively cooling the stator of the system, while previous designs required radiative cooling on the rotor. In 2004, a full-power integrated power and attitude control demonstration was executed demonstrating an operational speed range for these tests between $20,000$ and $60,000$ rpm and output bus voltages regulated at $125 {\rm V}$ during charge and discharge. A pair of momentum-locked counter-rotating flywheels can be used to store any excess energy that the main power source on-board the Icarus spacecraft produces, which could in turn be used to power attitude control systems.

\begin{figure}
\begin{center}
\includegraphics[width=12.15cm, height=5.31cm]{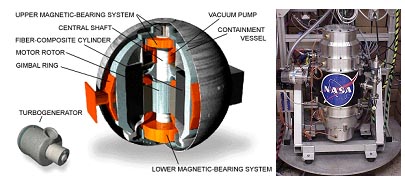}
\caption{Pairs of counter-rotating flywheels can be used as trickle-charged batteries. Their incredibly high angular momentum could be manipulated onto a robust attitude control module.} \label{fig3}
\end{center}
\end{figure}

It may seem unnecessary to use such low grade power sources, when a mission such as Icarus will contain enormous amounts of power through its main engine. However, the need for reliable backups, especially for essential subsystems such as star trackers, science and communications, make such considerations worth exploring if it would mean increasing the chances of success for the mission. An obvious way to power the spacecraft systems is by using a long life on board nuclear reactor. This could be used to initially power the fusion reactor and associated laser systems for fuel ignition but also other systems on board. 

The technology to power deep space missions has matured significantly since the time of the Voyager and Pioneer deep space probes. Radioisotope thermoelectric generators (RTGs), produce energy through thermocouples, where the heat source is a radioisotope such as Plutonium-238. Such devices have very long lifetimes, have no moving parts and require no maintenance but cannot be turned on and off and must be continuously cooled. Their lightweight and compact design, along with their high kW/kg ratio makes them ideal for interstellar missions. The contact flow of radiation means the rest of the instruments must be shielded from infrared interference. Although these devices are very expensive and require special governmental waivers to be used, they are currently the most reliable way of delivering power to a mission with a lifetime of more than 10 years.

The responsibility for coordinating all on board functions, as well as the nominal operation of the spacecraft is with the main computer system. However, each and every component of the spacecraft may suffer a partial or complete failure, a possibility which becomes increasingly likely considering the extremely long mission times, the important quantity here being the ``mean time between failures'' for each spacecraft component. There are several design considerations which can be made to increase the mission success rate.

There should be no single point failure, meaning there should be no one component which upon failure, compromises the main mission. This implies a level of redundancy to every system on the spacecraft, whether that is a computer, power, science or communication system. To reduce the number of redundant systems that must be taken along with the main spacecraft, the need for in-transit repair becomes obvious. However, this undertaking having its own engineering difficulties comes with yet another requirement, that there be no single point repair and that the computers are capable of fault isolation and containment. 

In many ways, this multiply redundant and self-checking approach aims to instil the computer with a fault-tolerant attitude, which in many ways resembles a personality. With this idea in mind, one can model the behaviour and reactions of the computer systems after biological systems, as a means of providing a framework from which a well-rounded design can spawn, as shown in Table 1.

For Icarus specifically, a distributive processing design approach has been envisioned. The spacecraft would contain identical computer subsystems simultaneously playing the role of an ``M out of N processor model'' redundant system, where M out of N systems must be in agreement for the decision to be executed. This architecture is conducive to a highly redundant design with uninterruptible, fault-tolerant system with multiple cross referenced secure copies of the mission profile. This way, the computer system is decentralized, with each computer being able to undertake any of the tasks necessary for the mission. The ``curiosity'' built into the main program can explore the surrounding medium, while the housekeeping commands continuously search for faults and effect repairs to damaged modules when necessary. A list of potential tasks or ``tickets'' is created and distributed throughout, which is then taken on by individual computer subsystems, when they are proven to be at the nominal operating condition and do not have more pressing things to do. The extensive modelling necessary to test the response of the Icarus computational and mechanical systems are a great contribution to the community in themselves. We intend on providing models for the operation of the full spacecraft and its subsystems, through which we intend on showing that we can achieve a high success rate. In any case, the models will be able to show where the mission design requires attention.

%\vspace{0.20in}
\vspace{0.20in}
\begin{table}[ht]
%\captionstyle{\centering}
\caption{\textit{The correspondence between artificial and biological behavioural responses, as a method for conceiving systems design and integration.}}
\vspace{0.20in}
\centering % used for centering table
\begin{tabular}{c c} % centered columns (4 columns)
\hline %inserts double horizontal lines
Artificial  & Biological \\ [0.5ex] % inserts table %heading
\hline % inserts single horizontal line
Self Repairing      &   Survival Instinct               \\
Autonomous Design      &   Independence/Individuality             \\
Vary Operational Goals      &   Choices/Curiosity          \\
Online/Offline Processing     &   Conscious/Subconscious          \\
Operational Overview of Subsystems/Probes      &   Maternal Instinct          \\

\hline %inserts single line
\end{tabular}
\label{table:qvac} % is used to refer this table in the text
\end{table}
\vspace{0.20in}

Once the computer programs had been designed and the mean time between failure models identified, a post-design aspiration would be to construct a computational tool that will execute all of the mission parameters with a high degree of accuracy and in real time. This undertaking has been named ``virtual Icarus'' or ``vIcarus'' and if taken on would comprise of a website containing all of the models and codes from the Icarus design study so as to enable a virtual launch of the Icarus spacecraft computationally and track its progress through every step of its mission in real time. A web interface would allow visitors to check in on the spacecraft, which would be feeding back detailed real time information on its trajectory and speed, as well as a list of the current command and execution protocols it is undertaking. If launched, vIcarus would be a full decades long mission to the selected target star and would represent a unique and historical undertaking with both pedagogical and scientific merits.

\begin{figure}
\begin{center}
\includegraphics[width=12.926cm, height=9.154cm]{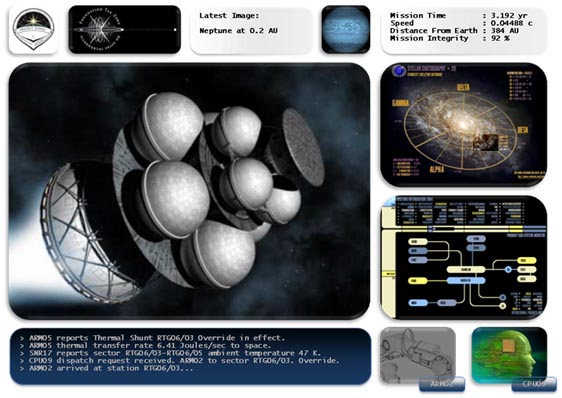}
\caption{An early imagining of the vIcarus web interface concept, summarizing mission statistics. The status of the spacecraft's subsystems could be explored and statistical information on the spacecraft would also be accumulated. } \label{fig4}
\end{center}
\end{figure}

Before all of the spacecraft systems are put in place however, the systems architecture has to be performed to provide a minimum baseline. This involves the launch and assembly of the Icarus vehicle. In the Daedalus study, it was assumed this architecture was in place. For Icarus, a number of systems architectures are evaluated - demonstrating what infrastructure options would need to be in place to support the assembly and launch of Icarus. A functional approach is adopted, where technologies rather than specific systems are identified initially, and from this, the architectures can be built. This then allows the evaluation of potential specific systems within the context of each architecture approach. The enabling technologies identified, will also share commonality with the Icarus vehicle itself, and undoubtedly, there will be potential for re-use in both directions. Enabling technologies identified may initially include:
\begin{itemize}
\item Robust Earth surface to Low Earth orbit transportation, and the options available based on national or commercial services, and the use of expendable, two-stage to orbit reusable and single stage to orbit reusable launch technology.
\item 	National or commercial interplanetary transportation and potential propulsion methods.
\item 	The use of propellant depots / stations.
\item 	Space dock construction facilities.
\item 	Local (in situ) resource utilisation.
\end{itemize}
The final design solution may be very similar to Daedalus or could even be radically different. In between there are various system and sub-system attributes which can also be compared to the Daedalus design. These are defined in the range of Options 1 - 6 as shown in Table 2. The characteristics will differ depending upon the system under study but the culmination of these different design solutions will lead to the pathways which allow the down selection to the final design. 

\vspace{0.20in}
\vspace{0.20in}
\begin{table}[ht]
%\captionstyle{\centering}
\caption{\textit{Design options for Icarus design solution.}}
\vspace{0.20in}
\centering % used for centering table
\begin{tabular}{c c} % centered columns (4 columns)
\hline %inserts double horizontal lines
Option  & Characteristics \\ [0.5ex] % inserts table %heading
\hline % inserts single horizontal line
1.0      &   Daedalus with improved calculations               \\
2.0      &   Daedalus with minor subsystem changes             \\
3.0      &   Daedalus with major system changes          \\
4.0     &   New design with major system attributes to Daedalus          \\
5.0      &   New design with minor subsystem attributes to Daedalus          \\
6.0      &   New and radically different design          \\

\hline %inserts single line
\end{tabular}
\label{table:qvac} % is used to refer this table in the text
\end{table}
\vspace{0.20in}

Icarus would represent a science driven mission and so some consideration would also have to be given to the type of instruments on board (e.g. magnetometers, spectrometers, radiometers…) as well as the science goals of the mission. The science goals would be split into a priority order. Primary science objectives would be along the lines of (1) Terrestrial planets (2) Giant planets (3) The star (4) Minor objects (5) Dust. Secondary science objectives would be along the lines of: observations of solar system outer bodies; measurements of the heliopause and interstellar medium; measurements addressing gravitational issues; spacecraft reliability with long duration missions. Table 3 shows some of the potential questions that an Icarus type probe could be designed to address as examples. However for the Icarus proposal the specific science objectives are yet to be determined. 

\vspace{0.20in}
\vspace{0.20in}
\begin{table}[ht]
%\captionstyle{\centering}
\caption{\textit{Potential science drivers.}}
\vspace{0.20in}
\centering % used for centering table
\begin{tabular}{l l} % centered columns (2 columns)
\hline %inserts double horizontal lines
Option  & Characteristics \\ [0.5ex] % inserts table %heading
\hline % inserts single horizontal line

Gravitation      &   What is the nature of the pioneer anomaly?     \\
                 &   What sources of gravitational waves can be detected?            \\

Heliosphere      &   What is the extent of the solar wind and its interaction \\
                 &   with the solar heliosphere?             \\

Planetary formation      &   What are the conditions for planet formation? \\
                         &   What is the extent of the habitable zone? \\     
                         &   How do other solar systems differ from ours? \\ 
                         
Stellar physics     &   What is the accuracy of long distance measurements to the stars       \\
                    &   What is the origin of low-frequency heliospheric radio emissions \\

Colonisation      &   Is human colonisation of the galaxy feasible?          \\
                  &   Can a technological species outlive its parent star?          \\

Life      &   Is their life on planets around other stars?          \\
          &   How long can life survive in deep space?          \\

Interstellar space      &   What is the mass function of objects in the Kuiper belt or Oort cloud?          \\
                        &   What are the properties of the interstellar medium?          \\
                        &   What is the abundance of interstellar nuclides?          \\
                        &   What are the properties of the interstellar medium?          \\
                        &   What is the cosmic ray background in interstellar space?          \\
                        &   What is the dust population resulting from collisions of Edgeworth-\\
                        &   Kuiper belt bodies?          \\

Solar system      &   Is our solar system typical in structure and metal content            \\
                  &   to others in the galaxy?                                             \\

Galactic     &   What is the age of the galaxy?          \\
             &   What is the nature of dark matter?          \\

Spacecraft      &   What is the long time survivability of a spacecraft           \\
                &   structure and electronics on long duration deep space missions?\\

\hline %inserts single line
\end{tabular}
\label{table:qvac} % is used to refer this table in the text
\end{table}
\vspace{0.20in}

\section{Project Icarus Programme Plans}

To address the project aims the research has been broken down into 20 modules (with sub-modules) which encompass all of the spacecraft systems and these are shown in Table 4. A programme of research is currently being planned to address all of these modules and this includes identifying the first trade studies. Some examples of potential early trade studies include: 
\begin{itemize}
\item	Finding out what the difference between projected state of the art in solar system observatories and Icarus carried or portable sensor capability might be in 2050 to confirm intuition that flyby is not useful.
\item	A study on flyby with decelerating probes vs. decelerating the whole spacecraft bus with the main propulsion system and then launching probes.  Trade space must include the communications system required to get the data back to the Solar System.  
\item	A review of historical solar system explorers that have been used to explore our own star system to assess what likely science drivers there will be from the science community. 
\item	A comparison between planetary acquisition of He$^3$ compared to accelerator generated He$^3$ (i.e. Deuteron $\rightarrow$ Deuterium target) and the likely cost difference.
\end{itemize}
The initial plan has been to set the project up in three stages: (1) establish initial design team and complete Terms of Reference by September of Year 1 (2) fully assemble design team (3) construct research programme by the beginning of Year 2 and Team Icarus officially begins technical work by the spring of Year 2. It has been estimated that with 20 volunteer designers Project Icarus will require around 35,000 total man hours spread over a five year research programme culminating in the final design. For comparison, Project Daedalus began on 10th January 1973 and the final reports were published 15th May 1978 taking just over 64 months or over 5 years. The study reports state that around 10,000 man hours were used by 13 core designers and several additional consultants [13]. For Project Icarus the design team will be split up into several layers as follows:
\begin{itemize}
\item	Core Design Team: This is the main design group that drives the project forward and does the majority of the work. All are personally known to each other and also manage the project.
\item	Floating Designers: These are designers who may not be personally known to the team but have agreed to contribute technically to the project by working on a system or sub-system.
\item	Consultants: These do not do any actual technical work but have mainly an advisory capacity to the project. 
\item	Reviewers: This is mainly made up of members of the Daedalus study group and has the function of providing a constructive technical review of any work produced by the team at various stages.
\end{itemize}
Project Icarus consists of 10 project phases with 5 key Stage Gates over the proposed five year duration. These are detailed in Table 5. An evaluation for entrance of each Stage Gate is based upon one of four metrics which judge the work of the previous phase: Go to proceed to the next phase; Kill as complete failure to meet Stage Gate so return to previous Stage Gate; Hold at current phase subject to further work due to current solution not being deemed acceptable; Recycle or repeat last phase where the work fails to demonstrate sufficient progress and the current solution is not acceptable. An assessment at each Stage Gate is based upon the determined deliverable such as the publication of a report.

\newpage

\vspace{0.20in}
\vspace{0.20in}
\begin{table}[ht]
%\captionstyle{\centering}
\caption{\textit{Research modules encompassing spacecraft systems.}}
\vspace{0.20in}
\flushleft % used for centering table
\begin{tabular}{l l} % centered columns (4 columns)
\hline %inserts double horizontal lines
Module 1-10 & Module 11-20 \\ [0.5ex] % inserts table %heading
\hline % inserts single horizontal line
1.0 Astronomical Target      &   11.0 Computing \& Data Management               \\
2.0 Mission Analysis \& Performance      &   12.0 Environment Control             \\
3.0 Vehicle Configuration      &   13.0 Ground Station \& Monitoring          \\
4.0 Primary Propulsion     &   14.0 Science          \\
5.0 Secondary Propulsion      &   15.0 Instruments \& Payload          \\
6.0 Fuel \& Fuel Acquisition      &   16.0 Mechanisms          \\
7.0 Structure \& Materials      &   17.0 Vehicle Assembly          \\
8.0 Power Systems      &   18.0 Vehicle Risk \& Repair          \\
9.0 Communications \& Telemetry      &   19.0 Design Realisation \& Technological Maturity        \\
10.0 Navigation \& Guidance Control      &   20.0 Design Certification          \\

\hline %inserts single line
\end{tabular}
\label{table:qvac} % is used to refer this table in the text
\end{table}
\vspace{0.20in}

%%%%%%%%%%%%%%
%%%%%%%%%%%%%%

\begin{table}[ht]
%\captionstyle{\centering}
\caption{\textit{Programme Plans (from the end of year 1).}}
\vspace{0.20in}
    \begin{tabular}{  l  p{6.5cm}  p{6.5cm} }
    \hline
Phase & Description & Summary \\ \hline
1 & Team assembly \& definition of terms of reference, by end of year 1. & Internal publication of Physics Requirements Document \& team assembly. This represents the official start of the project. \\ 
   
2 & Construction of work programme, up to spring of year 2. & Internal publication of Project Programme Document \& allocation of work programme, Stage Gate 1. \\
   
3 & Work programme conceptual design, by spring of year 3. & External publication of concept design options, Stage Gate 2. \\  
   
4 & Work programme preliminary design, by end of year 3. & Internal publication of preliminary design options. \\ 
   
5 & Preliminary design review, by early year 4. & Pass preliminary design review \& complete actions as appropriate, Stage Gate 3.  \\  
   
6 & Work programme, down select to detailed design options, by end of year 4. & Down select to Baseline Model \& internal publication of System Requirements Document.  \\ 
    
7 & Work programme, system integration, by summer of year 5. & Produce Integrated Baseline Model \& internal publication of Sub-System Requirements Document.  \\   
   
8 & Detailed design review, by end of summer of year 5 & Pass detailed design review and complete actions as appropriate, Stage Gate 4.   \\ 
   
9 & Certification of theoretical design solution, by end of year 5. & Internal publication of Icarus Certification Document, Stage Gate 5.    \\   
   
10 & Publication of final design solution, submit to JBIS by early year 6. & Publication of executive summary reports represents the key deliverable for Project Icarus.    \\ 
    
    \hline
    \end{tabular}
\end{table}
\vspace{0.20in}

%%%%%%%%%%%%%%%
%%%%%%%%%%%%%%%

Finally, one of the spin-off projects from this work is `Icarus Message in a bottle', the concept of carrying a message on board the spacecraft for any future generations. Note that the Pioneer and Voyager spacecraft have reached the outer reaches of our solar system and are carrying such a message. To contrast, the Pioneer plaque is about $0.12 \  {\rm kg}$ while the scientific payload of Daedalus was planned around 160 tons. It is not unreasonable to assume that the Icarus spacecraft could carry a several kilogram mass to contain such a message. This spin-off study is being undertaken within the framework of Faces from Earth (www.faces-from-earth.net) to design such a capsule. Designing and actually building the message would serve as an inspiring educational project for the general public, especially for the younger generation. 

\newpage

\section{Conclusions}

It has been over three decades since the landmark Daedalus engineering study and Project Icarus has finally arrived. This will be a complete redesign of the Daedalus systems including a re-examination of some of the original assumptions. An international team is now assembling to work on this exciting endeavour and bring the human dream of interstellar travel that much closer to being possible. Time will show if this ambitious project meets its overall goals in advancing fusion based space propulsion.

This work has been supported by the project consultants Marc Millis, Paul Gilster, Greg Matloff and Tibor Pacher in a personal capacity. Further information on Project Icarus is available from the online web site (www.icarusinterstellar.org). Since the submission of this paper, several new additions have joined the Project Icarus Study Group, some of which were present at the symposium. This includes Ian Crawford, Pat Galea, Adam Crowl, Robert Adams, Rob Swinney, Andreas Hein, Philip Reiss, Jardine-Barrington-Cook and James French. 
\\
\\
{\it Of Icarus, In ancient days two aviators procured to themselves wings. Daedalus flew safely through the middle air and was duly honoured on his landing. Icarus soared upwards to the sun till the wax melted which bound his wings and his flight ended in fiasco. The classical authorities tell us, of course, that he was only ``doing a stunt''; but I prefer to think of him as the man who brought to light a serious constructional defect in the flying-machines of his day. So, too, in science. Cautious Daedalus will apply his theories where he feels confident they will safely go; but by his excess of caution their hidden weaknesses remain undiscovered. Icarus will strain his theories to the breaking-point till the weak joints gape. For the mere adventure? Perhaps partly, this is human nature. But if he is destined not yet to reach the sun and solve finally the riddle of its construction, we may at least hope to learn from his journey some hints to build a better machine} \cite{Eddington}.

\section{Acknowledgements}

Team Icarus would like to thank the following members of the Daedalus study group for participating in the symposium discussions and giving support for this project. This includes Alan Bond, Bob Parkinson, Penny Wright, Jerry Webb, Geoff Richards and Tony Wright. Friedwardt Winterberg is also thanked for discussions. The staff of the British Interplanetary Society is thanked for hosting the successful symposium.


\begin{thebibliography}{99}

\bibitem{Schneider} J.~Schneider, {\it The extrasolar planets encyclopaedia},
http://exoplanet.eu/catalog.php, (date accessed November 2009).

\bibitem{Butler}
R.~Butler et al, {\it Catalog of nearby exoplanets}, Astrophys. J.,
\textbf{646}, pp505-522 (2006).

\bibitem{Udry}
S.~Udrey and N.~Santos {\it Statistical properties of exoplanets}, Ann. Rev. Astron. Astrophys.,
\textbf{45}, pp397 (2007).

\bibitem{Long1}
K.~Long, {\it Fusion, antimatter and the spacedrive: charting a path to the stars}, JBIS
\textbf{62}, No. 3, pp89-98 (2009).

\bibitem{Mallove}
E.~Mallove and G. ~Matloff, {\it The starflight handbook}, John Wiley and Sons, Inc, pp52 (1989).

\bibitem{Long2}
K.~Long, {\it The Status of the Warp Drive}, JBIS 
\textbf{61}, pp347-352 (2008).

\bibitem{Obousy}
R.~Obousy and G.~Cleaver, {\it Warp Drive: a New Approach}, JBIS 
\textbf{61}, pp364-369 (2008).

\bibitem{Forward1}
R.~Forward, {\it Antimatter Propulsion}, JBIS, 
\textbf{35}, pp391-395 (1982).

\bibitem{Forward2}
R.~Forward, {\it Interstellar travel using laser pushed light sails}, J. Spacecraft and Rockets \textbf{21}, pp187-195 (1984).

\bibitem{Taylor}
T.~Taylor, {\it Note on the possibility of nuclear propulsion of a very large vehicle at greater than earth escape velocities}, GAMD. \textbf{250}, (1957).

\bibitem{Dyson}
F.~Dyson, {\it Project Orion the atomic spaceship 1957-1965}, Penguin Books, USA (2002).

\bibitem{Winterberg}
F.~Winterberg, {\it Rocket propulsion by thermonuclear microbombs ignited with intense relativistic electron beams}, Raumfahrtforschung, \textbf{15}, pp208-217 (1971).

\bibitem{Bond}
A.~Bond and others, {\it Project Daedalus - the final report on the BIS starship study}, JBIS. (1978).

\bibitem{Parkinson}
B.~Parkinson, {\it Interplanetary, a history of the British Interplanetary Society}, BIS Production, (2008).

\bibitem{Orth}
D.~Orth, {\it Parameter studies for the VISTA spacecraft concept}, UCRL-JC \textbf{141513}, (2000).

\bibitem{Beals}
K.~Beals et al, {\it Project LONGSHOT, an unmanned probe to Alpha Centauri}, N89-16904, (1988).

\bibitem{Mankins}
J.~Mankins, {\it Technology readiness levels a white paper}, PNASA Internal White Paper, (1995).

\bibitem{Schulze}
N.~Schulze, {\it Fusion energy for space missions in the 21st century}, NASA TM4297/4298, (1991).

\bibitem{Pfalzner}
S.~Pfalzner, {\it An introduction to inertial confinement fusion}, Taylor \& Francis, (2006).

\bibitem{Nuckolls}
J.~Nuckolls et al, {\it Laser compression of matter to super-high densities: thermonuclear (CTR) applications}, Nature \textbf{239}, pp129 (1972).

\bibitem{Lindl}
J.~Lindl, {\it Inertial confinement fusion, the quest for ignition and energy gain using indirect drive}, Springer, (1998).

\bibitem{Pick}
M.~Pick, {\it technological achievements and experience at JET}, JET-P(98)34, Presented  20th symposium on fusion technology, France, December 1998.

\bibitem{ITER}
{\it ITER the design phase, http://www.iter.org/pdfs/ITER\_Design\_Phase.pdf} (accessed November 2009).

\bibitem{Hammel}
B.~Hammel et al, {\it The NIF ignition program: progress and planning}, Plasma Phys.Control.Fusion 48, \textbf{B497-B506}, (2006). 

\bibitem{Fleurot}
N.~Fleurot et al, {\it The Laser Megajoule (LMJ) project dedicated to inertial confinement fusion: development and construction status}, Fusion Engineering \& Design, \textbf{V74}, pp147-154, (2005). 

\bibitem{Atzeni}
S.~Atzeni, {\it Fast ignitor target studies for the HiPER project}, Physics of Plasmas, \textbf{15}, 056311, (2008).

\bibitem{Leger}
A.~Leger, {\it Transiting exoplanets from the CoRoT space mission},Astron. Astrophys., \textbf{506}, 287-302, (2009).

\bibitem{YEE}
J. ~Yee and S. ~Gaudi {\it Characterizing long-period transiting planets observed by Kepler}, Astrophys. J., \textbf{688}, 616-627 (2008).

\bibitem{Lawson}
P. ~Lawson et al {\it Terrestrial planet finder interferometer science working group report}, NASA JPL publication 07-1, March 2007.

\bibitem{Cockell}
S.~Cockell, {\it Darwin - a  mission to detect and search for life on extrasolar planets}, Astrobiology,  \textbf{9}, pp1-22 (2009).

\bibitem{Gatewood}
G.~Gatewood, {\it A study of the astrometric motion of Barnard's star}, Astrophys. Space Sci.,  \textbf{223}, pp91-98 (1995).

\bibitem{Hatzes}
A.~Hatzes, {\it Evidence for a long-period planet orbiting $\epsilon$ Eridani}, Astrophys. J.,  \textbf{L145-L148}, (2000).

\bibitem{Quillen}
A.~Quillen and S. ~Thorndike, {\it Structure in the $\epsilon$ Eridani dusty disk caused by mean motion resonances with a 0.3 eccentricity planet at periastron}, Astrophys. J.,  \textbf{L149-L152}, (2002).

\bibitem{Backman}
D.~Backman, {\it Epsilon Eridani's planetary debris disk: structure and dynamics based on Spitzer and Caltech submillimeter observations}, Astrophys. J., \textbf{690}, pp1522-1538 (2009).

\bibitem{Jones}
B.~Jones, D.~Underwood and P.~Sleep {\it Prospects for habitable 'Earths' in known exoplanetary systems}, Astrophys. J., \textbf{622}, pp1091-1011 (2005).

\bibitem{Greaves}
J.S~Greaves, M.C.~Wyatt W.S.~Holland and W.R.F. Dent {\it The debris disc around Tau Ceti: a massive analogue to the Kuiper Belt}, MNRAS., \textbf{351}, L54-L58 (2004).

\bibitem{Eggenberger}
A.~Eggenberger, S.~Udry and M.~Mayor, {\it Statistical properties of exoplanets III. Planet properties and stellar multiplicity}, Astron. Astrophys.,    \textbf{417}, pp353-360 (2004).

\bibitem{Barbieri}
M.~Barbieri, F.~Marzari and H.~Scholl {\it Formation of terrestrial planets in close binary systems: The case of $\alpha$ Centauri A}, Astron. Astrophys., \textbf{396}, pp219-224 (2002).

\bibitem{Quintana}
E.~Quintana et al, {\it Terrestrial planet formation in the $\alpha$ Centauri system}, Astrophys. J., \textbf{576}, pp982-996 (2002).

\bibitem{Scalo}
J.~Scalo  et al, {\it M stars as targets for terrestrial exoplanet searches and bio signature detection}, Astrobiology, \textbf{7}, pp85-166 (2006).

\bibitem{Lewis}
J.~Lewis, {\it Mining the sky, untold riches from the asteroids, comets and planets}, Helix books, (1996).

\bibitem{Kammash}
T.~Kammash, {\it Antiproton driven magnetically insulated inertial confinement fusion (MICF) propulsion systems}, NIAC 98-02 final report. 

\bibitem{Perkins}
L.~Perkins  and J ~Hammer, {\it Advanced Fusion Concepts: Antiproton Catalysis}, June 1993.

\bibitem{Frolov} 
A. ~Frolov  and A.~Thakkar, {\it Weakly bound ground states in three-body Coulomb systems with unit charges}, Phys. Rev \textbf{A}, 46 (1992).

\bibitem{Kammash1}
T.~Kammash et al, {\it Antimatter Driven $P-B^{11}$ AIP Conference Proceedings} \textbf{V654}, pp497-501 (2003).

\bibitem{Zubrin}
R.~Zubrin and D.~Andrews, {\it Magnetic sails and interplanetary travel}, AIAA, 89-2441, AIAA/ASMA 25th Joint Propulsion Conference, Monterey, Calif., July 10-12 (1989).

\bibitem{Patel}
R.~Patel, {\it Spacecraft power systems}, CRC Press, (2004).

\bibitem{Eddington}
A.~Eddington, {\it Stars \& Atoms}, Oxford University Press, pp41,  (1927).

\end{thebibliography}
\end{document}